\documentclass[ reprint, amsmath,amssymb, aps]{revtex4-1}
\usepackage{graphicx}
\usepackage{dcolumn}
\usepackage{bm}
\usepackage{amssymb}
\usepackage{amsmath}
\usepackage{epsf}
\usepackage{color}
\usepackage{gensymb}
\usepackage{natbib}
\draft

\begin{document}

\preprint{APS/123-QED}
\title{Phase coexistence and associated non-equilibrium dynamics under simultaneously applied magnetic field and pressure\\}

\author{Sudip Pal, Kranti Kumar and A. Banerjee}
\affiliation{%
 UGC DAE Consortium for Scientific Research\\
 Khandwa road, University campus\\
 Indore-452001, M.P., India 
}%
\begin{abstract}
 A quantitative estimation of the effect of simultaneously applied external pressure (P) and magnetic field (H) on the phase coexistence has been presented for Pr$_{0.5}$Ca$_{0.5}$Mn$_{0.975}$Al$_{0.025}$O$_{3}$ and La$_{0.5}$Ca$_{0.5}$MnO$_3$, where the ferromagnetic (FM)-metal and antiferromagnetic (AFM)-insulator phases compete in real space. We found that the nonequilibrium dynamics across the FM-AFM transition is primarily dictated by the effect of P and H on the supercooling, superheating temperatures, and the nucleation and growth rate of the equilibrium phase. These effects across the transition is also responsible for the relative volume fraction of the competing phases at low temperature. Importantly in the entire magnetic field-pressure-temperature range of phase coexistence, the interface between the two competing phases having different spin and structural order plays a very important role in controlling the non-equilibrium dynamics.
\end{abstract}                            
\maketitle
\section{Introduction:} 
The coexistence of competing phases in spatially separated regions is a common phenomenon across first order phase transitions. Such coexistence has been studied in the context of various class of promising  materials, for instance, multiferroics showing strong magnetodielectric coupling,  Mott insulators, artificial spin ice etc. and of course manganites, shape memory alloys showing colossal magneto-resistance, magnetocaloric effects {\color{blue}\cite{Dagotto2002, Ahn2004,SBR2004,Dagotto2005,Lin2018,Zhu2016}}. The increasing research interest in this direction is driven firstly, due to a possibility of technological application arising from tunability of contrasting phases and secondly, these phenomena are not yet fully understood. Control over the volume fraction of the competing phases and identification of the critical parameters are required to rein in their physical properties {\color{blue}\cite{Zang2015,Pandey2003,Lourembam2014,Song2021}} and utilize them in spin valves, memory devices etc {\color{blue}\cite{Yang2019}}. In this context, besides magnetic and electric fields, external pressure can be a very good control parameter due to strong involvement of the underlying lattice in these systems, which has been recently exemplified using strain in La$_{0.7}$Ca$_{0.3}$MnO$_3$ {\color{blue}\cite{Hong2020}}. In recent times, advancement in experimental techniques have turned pressure (P) into an additional, as well as important state-of-the-art experimental tool. However, investigation involving the variation of both magnetic field (H) and P along with necessary quantitative understanding is still missing. In this backdrop, manganites can serve as an excellent prototype {\color{blue}\cite{Hong2020,Yang2019,Baldini2012,McLeod2017,Joseph2019,Hong2018,Udalov2020,Zhou2020}}.

Colossal magnetoresistance in Manganites, large magnetocaloric and magnetostriction in various alloys and intermetallics are directly related to the coexistence of metallic-ferromagnetic (FM) and insulating-antiferromagnetic (AFM) phases appearing due to the first order magnetic transition (FOMT) {\color{blue}\cite{SBR2013, Markovich2017,Banik2018}}. In our study, we have combined both H and P to tune and determine the origin of the tunability of the phase coexistence in two well known manganites Pr$_{0.5}$Ca$_{0.5}$Mn$_{0.975}$Al$_{0.025}$O$_{3}$ (PCMAO) and La$_{0.5}$Ca$_{0.5}$MnO$_3$ (LCMO). Our aim is twofold:  How simultaneous application of H and P influence the phase coexistence across the first order transition i.e., across the H-T region where magnetization exhibits thermal hysteresis due to supercooling and superheating phenomena. It is interesting to note further that PCMAO and LCMO show phase coexistence even below the thermal hysteresis which persists down to lowest temperature due to kinetic arrest of the first order phase transition. Therefore, we are also interested on the effect of P and H at the low temperature regime below the hysteresis. Throughout the study, we emphasize on an important fact that in both of these regions (across and below the thermal hysteresis), the phase coexistence is a nonequilibrium phenomenon and the time scale associated with the non-equilibrium dynamics can be tuned by H and P. In addition, a detail quantitative understanding of the volume fraction of the coexisting phases is necessary to exploit the phase coexistence as a device. Here, we have used bulk magnetization value to estimate the volume fraction of the FM-metal and AFM-insulator phases. It may be noted that, until now mostly the surface sensitive techniques, like Hall probe, scanning tunneling microscopy, magnetic force microscopy, photo electron spectroscopy etc. have been used as a probe of phase coexistence {\color{blue}\cite{SBR2004,Dagotto2005,Lin2018,Zhu2016}}, but the surface often contains additional disorder and strain. As a result, the nature of the phase coexistence in the bulk may be different.

A first order transition can be distinguished from a continuous transition by the appearance of supercooling (H$^*$, T$^*$)  and superheating (H$^{**}$, T$^{**}$)  lines which give rise to a hysteresis across the transition in the H-T plane. However, quenched disorder present in the sample broadens these lines into bands {\color{blue}\cite{Imry1979}} and results into the phase coexistence in a wide range of the control parameters. In several materials, the kinetics of a FOMT gets arrested below a heuristic band, called kinetic arrest band (H$_K$, T$_K$).  The transformation from the high-T phase to low-T phase is arrested below that part of the [H$_K$, T$_K$] band which lies above the [H$^*$, T$^*$] band and the system below the thermal hysteresis remains as a composite of the transformed fraction of the low-T phase and untransformed fraction of the high-T phase {\color{blue}\cite{Chattopadhyay2005, AB2006,AB2008,Chaddah2008,Dash2010, AB2011, EPL2013,Sudip2021,Pal2021}}. The phenomena is called as the ``kinetic arrest'' (KA) and the magnetic state at low temperature is called as the magnetic glass. In this state, the high-T phase persists down to lowest temperature because of KA, and the phenomenon of KA provides a systematic framework to understand the bizarre properties of phase separation in many systems. This phenomenon has been observed across various transitions, including magnetic, structural, metal-insulator transitions, in multiferroics etc {\color{blue}\cite{SWC2006,Choi2010,MnFePSi,Mn2PtGa,Cakir2020,Pampi2020}}.

PCMAO is in the paramagnetic state at room temperature and undergoes transition to CE type antiferromagnetic (AFM) phase with decrease in temperature {\color{blue}\cite{AB2006}}. The high-T AFM state is eventually followed by a FOMT to low-T FM phase on further cooling. At low strength of the magnetic field, the AFM to FM phase transition is partially arrested during cooling and the system contains a large fraction of the AFM phase down to lowest temperature {\color{blue}\cite{AB2006}}. On the other hand, in LCMO, it undergoes a first order transition from high-T FM to low-T CE type AFM phase which is arrested at high fields {\color{blue}\cite{AB2008}}. The arrested FM fraction increases with increase in H.  Present work includes the magnetization measurements under simultaneous application of P and H and a quantitative investigation on the effects of H and P on the volume fraction and the non-equilibrium dynamics of the competing phases associated with the first order transition. We explain the observation in terms of the interplay of H and P induced changes in the supercooling and superheating across the first order transition and the kinetic arrest. We also underline that the interface between the two competing phases plays a crucial role in controlling the dynamics of the phase coexisting state. 
\section{Experimental details:} 
Polycrystalline bulk samples of PCMAO and LCMO have been prepared by standard solid state method and chemical route, known as pyrophoric method, respectively {\color{blue}\cite{Chaddah2008, AB2006}}. Further details on the sample preparation and characterization can be found in Ref. {\color{blue}24} and {\color{blue}26}. Magnetic measurements have been carried out in commercial 7 Tesla superconducting quantum interference device (SQUID) magnetometer (M/S Quantum Design, USA). For the measurements under pressure, a Cu-Be cell (Mcell 10, easy lab) has been used, where a fixed pressure was applied and locked at room temperature. The pressure cell can be mounted inside the magnetometer to carry out the magnetization measurements, which have been performed at different H at a fixed pressure. Daphne oil has been used as the pressure transmitting medium to maintain the hydrostatic condition of pressure on the sample. Pressure values have been determined from the variation of  superconducting transition temperature of Sn wire that has been loaded inside the pressure cell {\color{blue}\cite{Dash2010}}. Field cooled cooling (FCC) magnetization (M) versus T curves are recorded while cooling the sample is presence of a constant magnetic field. The field cooled warming (FCW) curves are recorded in the subsequent warming cycle without changing the field. For zero field cooled (ZFC) measurements, the sample has been cooled down to T = 5 K in absence of external field. Then, M versus T data has been recorded during warming after applying external field at the lowest temperature. Time ($t$) dependent magnetization (M) data have been recorded in field cooling (FC) mode where the sample is first cooled from T = 320 K to the measurement temperature in presence of the external field. After temperature becomes stable, then isothermal variation of M with time recorded keeping the field unchanged.
\begin{figure}[t]
\centering
\hspace{-0.8cm}
\includegraphics[scale=0.30]{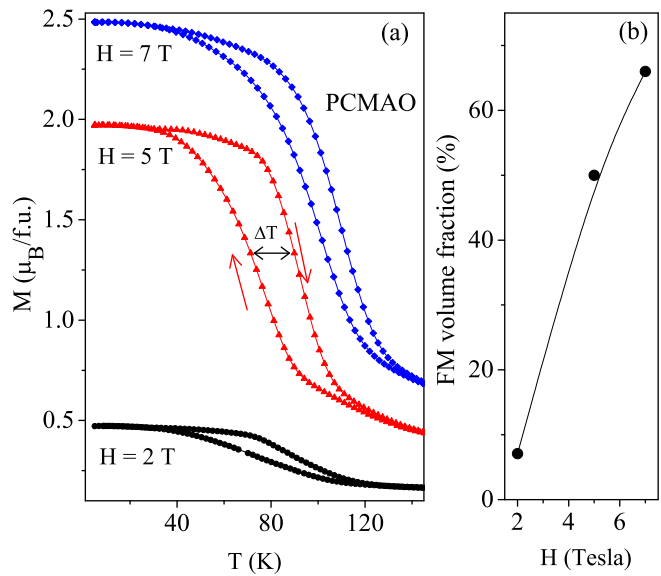}
\caption{(a) Temperature variation of magnetization of PCMAO at H = 2, 5 and 7 Tesla recorded in FCC and FCW modes. The thermal hysteresis indicates the transition between high-T AFM and low-T FM phases and it shifts towards higher temperature with magnetic field. $\Delta$T is the width of the thermal hysteresis. (b) Variation of FM volume fraction ($f_{FM}$) with H at T = 5 K. The line is a guide to eye.} 
\end{figure}
\begin{figure}[t]
\centering
\hspace{-0.8cm}
\includegraphics[scale=0.42]{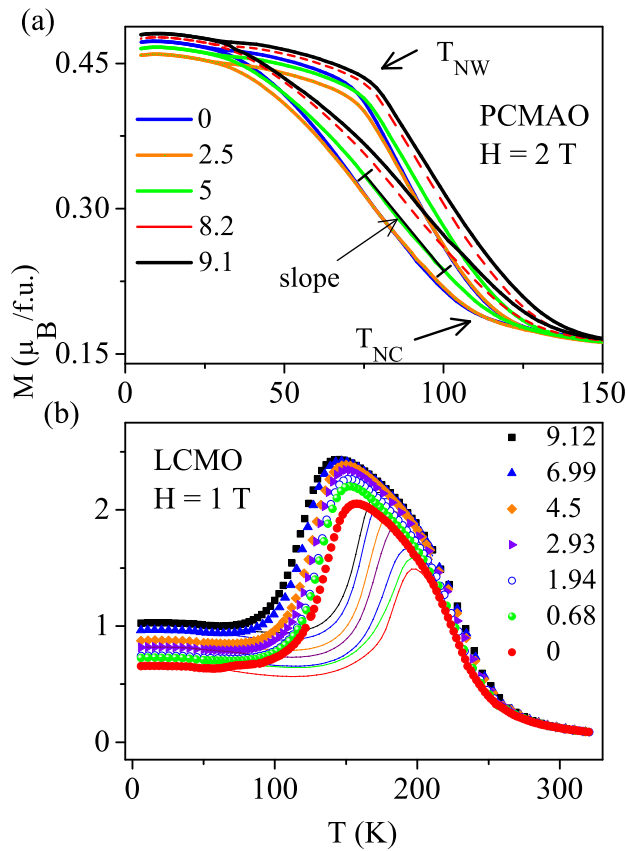}
\caption{Temperature variation of magnetization in different pressure of (a) PCMAO at H = 2 Tesla. In the FCC curve at P = 5 kbar, we have drawn a straight line to determine the slope of the curve, which should be propertional to the transformation rate from high-T AFM to low-T FM phase across the thermal hysteresis. (b) LCMO at H = 1 Tesla. The legends indicate the P values applied on the sample in the unit of kbar. In case of LCMO also, the transformation rate from high-T FM to low-T AFM phase can be found from the slope of the FCC curves at respected pressure values.} 
\end{figure}

\section{Results:}
Fig. {\color{blue}1(a)} presents the temperature variation of M (FCC and FCW) across the transition between high-T AFM and low-T FM phases in PCMAO at three different magnetic fields. Magnetization shows thermal hysteresis across the transition and the hysteresis shifts towards higher temperature with increase in H. Also note that, the width of the thermal hysteresis ($\Delta$T, see Fig. {\color{blue}1(a)}) varies non-monotonically with H. At H = 5 Tesla, the width $\Delta$T $\approx$ 20 K, which is significantly larger than $\Delta$T $\approx$ 12 and 10 K at H = 2 and 7 Tesla respectively. On the other hand, the value of M at low temperatures, say at T = 5 K increases by a large amount with increase in H which can only be explained with the phase separation scenario. It should be mentioned here that the saturation field of the FM phase in PCMAO is a few hundred Oe {\color{blue}\cite{AB2006}}. At low temperature, the system contains both FM and AFM volume fractions, and the amount of the FM phase increases with H at the expense of the competing AFM phase {\color{blue}\cite{AB2006}}. In Fig. {\color{blue}1(b)}, we have estimated the variation of the FM phase fraction at T = 5 K with the applied magnetic field. It has been estimated from the magnetization value at T = 5 K ($M_{5K}$) following the method described in Ref. {\color{blue}30}. If the FM volume fraction is $f_{FM}$ at T = 5 K [so the AFM volume fraction is (1-$f_{FM}$)], then $M_{5K}$ is given by the relation-
\begin{equation}
M_{5K} = f_{FM} (M_0 + \alpha \times H_m) + (1 - f_{FM}) \times \beta \times H_m
\end{equation}
Here, $\alpha$ and $\beta$ are the dc susceptibility (dM/dH) of the FM and AFM phases respectively which are obtained from the slope of the M-H curve at T = 5 K. $M_0$ is the spontaneous magnetization of the FM phase. As H increases from H = 2 to 7 Tesla, $f_{FM}$ also increases by more than eight folds from around 7 to 60 \%.\\ 
In Figs. {\color{blue}2(a)} and {\color{blue}2(b)}, we have shown the FCC and FCW curves of PCMAO and LCMO respectively. Different curves correspond to the different applied pressure at a constant H = 2 and 1 Tesla for PCMAO and LCMO respectively. It may be mentioned here that a part of this data of LCMO has been reported earlier in Ref. {\color{blue}28} but has been reproduced here to make this paper self content.  The broad thermal hysteresis between the FCC and FCW curves evince the coexistence of the high and low temperature phases across the hysteresis. Besides, due to kinetic arrest, the phase coexistence prevails down to the lowest temperature in both these systems. In PCMAO, the M at T = 5 K ($M_{5K}$) shows non-monotonic variation with pressure. It initially decreases as we increase the pressure from ambient to P = 2.5 kbar. As pressure increases further, $M_{5 K}$ starts to increase again. On the other hand in LCMO, $M_{5 K}$ monotonically increases with pressure. Such variation of $M_{5 K}$ indicates that the volume fraction of the two coexisting phases (FM and AFM) change with pressure in both systems, but the dependence is qualitatively very much different. In the next section, we will estimate the volume fraction of the FM phase at T = 5 K in H-P landscape and show that the efficiency of pressure in tuning the volume fraction of the coexisting phases also depends upon the applied magnetic field.
\begin{figure}[t]
\centering
\includegraphics[scale=0.42]{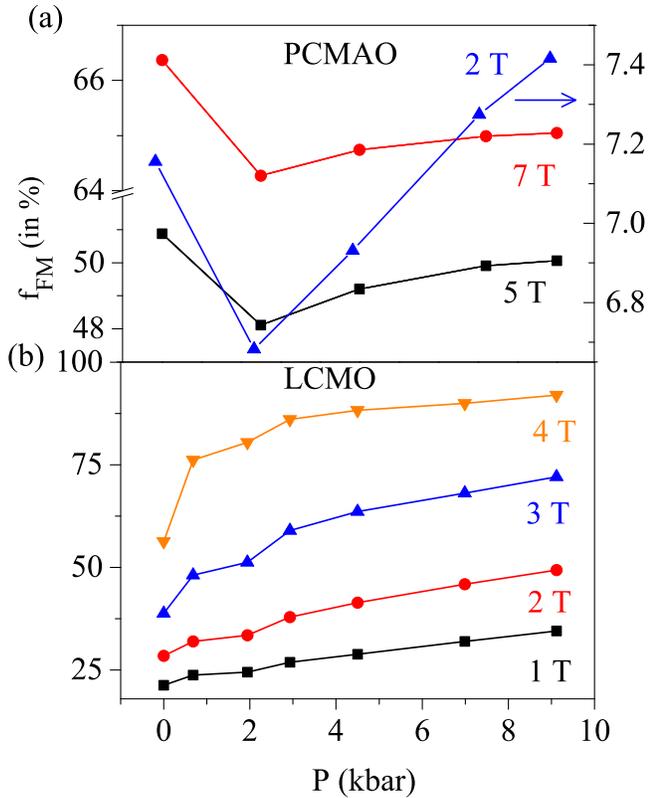}
\caption{ Change in volume fraction of the metallic-FM phase ($f_{FM}$) with H and P in (a) PCMAO and (b) LCMO. Note that in case of PCMAO, $f_{FM}$ initially decreases and finally starts to increase again with pressure. On the other hand in LCMO, $f_{FM}$ monotonically increases.} 
\end{figure}
\subsection{Effect of pressure and magnetic field on the FM and AFM phase fractions at low temperature:}  

In Figs. {\color{blue}3(a)} and {\color{blue}3(b)}, we present the variation of the FM volume fraction ($f_{FM}$ in \%) at T = 5 K in PCMAO and LCMO, respectively at different P and H, which we have calculated using eqn. {\color{blue}1}. It is interesting that in case of PCMAO [Fig. {\color{blue}3(a)}], pressure initially suppresses the FM phase fraction. At higher pressure, $f_{FM}$ increases again. Moreover, the change in phase fraction with the application of pressure is small. For example (see H = 5 Tesla, Fig. 3(a), black filled square), at H = 5 Tesla $f_{FM}$ is around 50.8\% at ambient pressure which is consistent with Fig. {\color{blue}1(b)}. As pressure is increased to P = 2.5 kbar, $f_{FM}$ reduces to $f_{FM}$ = 48\% which is followed by monotonic rise to 50.1\% at P = 9.1 kbar. Such kind of non-monotonic dependence of $f_{FM}$ on pressure is interesting. It points towards the existence of two competing mechanisms occurring simultaneously, one trying to enhance the $f_{FM}$ and the other mechanism tending to reduce the $f_{FM}$ at each pressure. As we will show later that this is actually the situation and the net volume fraction of the two phases is decided by which of these two mechanism is dominant.  On the other hand, in case of LCMO as shown in Fig. {\color{blue}3(b)}, the change in the FM volume fraction is comparatively large and it monotonically increases with increase in the pressure. For example, at H = 1 Tesla, $f_{FM}$ = 21.1\% at ambient pressure which monotonically rises to 34.5\% at P = 9.12 kbar. Also note that at low fields say H = 1 Tesla, $f_{FM}$ varies rather smoothly as a function of pressure.  However at higher H say 4 Tesla, $f_{FM}$ initially increases sharply and tends to saturate with increase in the applied pressure. The variation of $f_{FM}$ in both PCMAO and LCMO reveal a complex dependence of the volume fraction of the competing metallic-FM and insulating-AFM phases on H and P.
\begin{figure}[t]
\centering
\hspace {1.5 cm}
\includegraphics[scale=0.36]{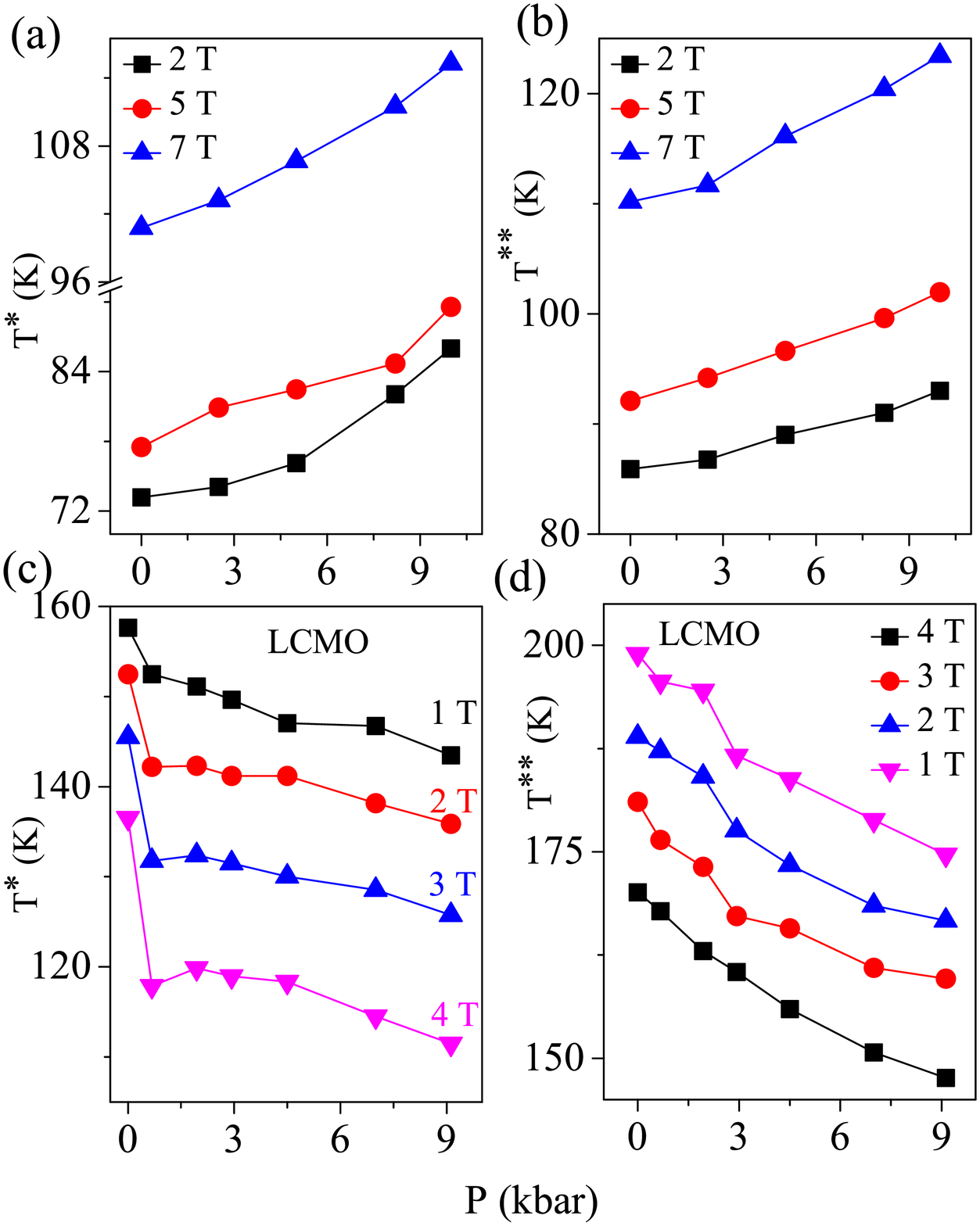}
\caption{(a) and (b) shows the variation of the T$^{*}$ and T$^{**}$s with P at different applied field in PCMAO. (c) and (d) shows the same in case of LCMO. The solid lines are guide to eye.} 
\end{figure}
\subsection{Effect of pressure on the supercooling, superheating and kinetic arrest temperatures:}
Thermal hysteresis in PCMAO progressively shifts toward higher temperature as P increases [see Fig. {\color{blue}2(a)}] i.e., the supercooling (T$^*$) and superheating (T$^{**}$) temperatures increase with P. We want to mention here that we have performed magnetization measurements at different P and H, but have not shown here for conciseness. Some data on LCMO can be found in Ref. {\color{blue}28}. The variation of the T$^*$ and T$^{**}$ with H and P are shown in Figs. {\color{blue}4(a)} and {\color{blue}4(b)}. There are different ways to define T$^*$ and T$^{**}$ because the thermal hysteresis is broad. In case of PCMAO, we have considered  the temperature where the (dM/dT) in FCC (FCW) shows the minima as T$^*$ (T$^{**}$).   In the FCC curve, M start to increase rapidly at T$_{NC}$, which indicates the rapid nucleation and growth of the FM phase below T$_{NC}$. Similarly, in the FCW curve, nucleation of the AFM phase has started above T$_{NW}$. Interestingly note that,  T$_{NC}$ $>$ T$_{NW}$, which means that onset of the nucleation of the AFM phase during heating has started at a lower temperature than the onset temperature of nucleation of FM phase during cooling. This indicates towards a wide distribution of the T$^*$ and T$^{**}$ in the material, such that they form bands instead of lines in H-T plane, and overlap with each other. This occurs due to presence of quenched disorder, which results into a landscape of the transition temperature, and thereby broadening of the first order transition {\color{blue}\cite{SBR2004,Imry1979}}. In LCMO, the thermal hysteresis shifts towards lower temperature as higher P is applied. In this case, we have taken a different approach to find out the T$^*$ and T$^{**}$ values. In the FCC curve, the temperature where M shows peak i.e., M is maximum is taken as the T$^*$. Similarly, T$^{**}$ has been taken where M has a peak in the FCW curve. We have shown the variation of T$^*$ and T$^{**}$ in Figs. {\color{blue}4(c)} and {\color{blue}4(d)}. The effect on T$^{*}$ by P and H is quite dramatic in case of LCMO. T$^{*}$ falls sharply at very tiny pressure of P = 0.68 kbar (Fig. {\color{blue}4(c)}, remains comparatively unaffected at higher pressure. It is important to note that, there are a few common features if we compare the effect of pressure on PCMAO and LCMO. First of all, application of P suppresses the AFM phase in both the systems, which is evident from the variation of T$^*$ and T$^{**}$ with pressure. Secondly, the variation of these transition temperatures are not linear in P. \\

Besides, T$^*$ and T$^{**}$, the temperature T$_K$ corresponding to kinetic arrest band is also important in case of both PCMAO and LCMO. The phase coexistence persist even below the thermal hysteresis down to lowest temperature due to kinetic arrest band, [H$_K$, T$_K$], and therefore, it is necessary to check the position of the [H$_K$, T$_K$]  band at higher pressure. In case of PCMAO, the effect of P on the [H$_K$, T$_K$] band at a fixed H can be found from the measurement of the ZFC curves as shown in the Fig. {\color{blue}5} {\color{blue}\cite{Lakhani2010}}. In the case of zero field cooling, as T is increased, magnetization sharply increases with temperature due to transformation of arrested AFM phase into the FM phase-which is called dearrest. M increases up to the end point of  [H$_K$, T$_K$] band, where the value of M is maximum, which we have marked as vertical line in Fig. {\color{blue}5}. Note that, this temperature is not significantly affected by pressure. Above this temperature, no significant conversion of the phase fraction happens until [H$^{**}$, T$^{**}$] band is reached, where the FM phase converts into AFM phase and M sharply decreases. We have also indicated the T$^{**}$ values by the downward arrows for the ambient and highest applied pressures. This data reveals that the external pressure does not affect the [H$_K$, T$_K$] band significantly as compared to the [H$^{**}$, T$^{**}$] band. It should be mentioned here that [H$_K$, T$_K$] is practically a band, i.e. it is spread over certain H-T window and the ZFC curve as shown Fig. {\color{blue}5} only gives one end of the band. Nonetheless, it can be shown that pressure does not significantly affect the width of the band as well (data no shown here). In case of LCMO also, P does not affect the [H$_K$, T$_K$] band significantly. The data has not been shown here for conciseness.
\begin{figure}[t]
\centering
\includegraphics[scale=0.36]{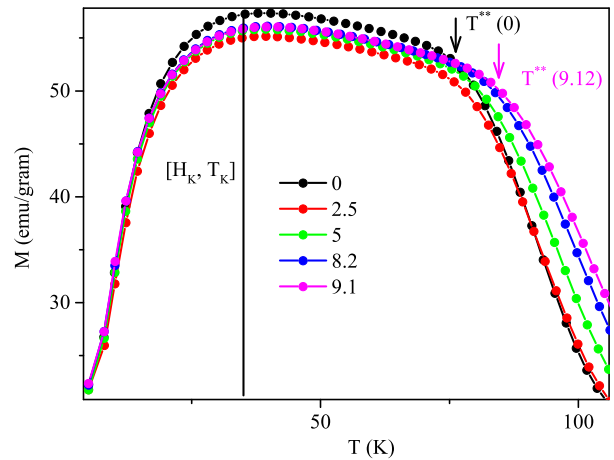}
\caption{ZFC curves of PCMAO at different P. The measurement H = 5 Tesla. The vertical line marks the end of KA band. It indicates that [H$_K$, T$_K$] remains unaffected by P. } 
\end{figure}
\subsection{Effect of pressure on the nucleation and growth across the thermal hysteresis: } 
A first order transition during cooling proceeds through the nucleation and subsequent growth of the low-T phase at the cost of the high-T parent phase. Therefore, in case of PCMAO, as we gradually reduce the sample temperature, the high-T AFM phase transforms into the FM phase through the creation and growth of the FM droplets which are larger than a critical size {\color{blue}\cite{Rawat2013}}. This is manifested as the increase in M with decrease in temperature across the thermal hysteresis shown in Figs. {\color{blue}1(a)} and {\color{blue}2(a)}. Therefore, legitimately, the transformation rate of the FM phase should be proportional to the slope of the M-T curve in the hysteresis region. The slope of the FCC curves shown in Fig. {\color{blue}2(a)} in the hysteresis region decreases with the increase in P. The decrease in the slope indicates that the transformation rate of the FM phase declines with increasing P at a fixed temperature and field. To further confirm this, we have measured the isothermal relaxation of M with time at a fixed T in the hysteresis region, but at different P.  We have recorded the time dependence of M in the field cooled state, i.e. the sample has been cooled to the probing temperature in presence of the field and after stabilizing the temperature, magnetization has been recorded for next few hours at the same field. The data has been shown in Fig. {\color{blue}6(a)} for T = 65 K at H = 4 Tesla. First of all, note that M increases with time, which confirms that high-T AFM phase transforms into the FM phase during cooling across the hysteresis region, which is expected. Now the important observation is, as P increases the relaxation rate, i.e. how fast M changes with time, becomes slower, which indicates that the growth rate of the FM phase fraction decreases at higher pressure. This is consistent with the fact that the slope of the M-T curve decreases with P. Similarly, in LCMO, the transition during cooling proceeds through the transformation of the high-T FM phase into the low-T AFM phase. It can be shown that the slope of the M-T curves in Fig. {\color{blue}2(b)} decreases with increasing P, which indicates towards the suppression of the growth of the AFM phase with increase in the pressure. In Fig. {\color{blue}6(b)}, we have also shown the time dependence of M after the sample has been cooled in the field cooled protocol to T = 100 K in presence of H = 6 Tesla, which is within the thermal hysteresis region. In this case also, the transformation rate reduces at  higher pressure, and hence conforms that P suppresses the transformation rate from FM to AFM phase across the thermal hysteresis. Now, note that  these observations are consistent with the fact that pressure favors the FM phase in both the systems, which we have concluded in the last section (Fig. {\color{blue}4}). 

Now, we will probe the effect of pressure on the metastable behaviour across the entire thermal hysteresis region in case of LCMO. In Fig. {\color{blue}6(c)}, we have shown the time dependence of M (relaxation) at different temperatures within T = 100 to 40 K at ambient pressure and H = 6 Tesla which is the thermal hysteresis region. To measure these relaxation curves, we have cooled the sample from room temperature (i.e. paramagnetic state) to the measurement temperatures in presence of the H = 6 Tesla magnetic field. After the temperature becomes stable, magnetization has been recorded for next few hours without changing the field. First of all note that, M decreases with time at all the temperatures because the high-T FM phase transforms into the AFM phase. Now, the change of M with time i.e. the relaxation rate initially increases as we reduce the temperature from 100 K. For example, the relaxation at T = 80 K is faster than at 100 K. However, the relaxation again decreases with further decrease in the temperature down to 40 K. The increase in the relaxation rate with decreasing temperature can be understood in the framework of kinetically arrested first order transition {\color{blue}\cite{Chaddah2008,Pal2021}}. As temperature reduces, the free energy barrier between the high and low-T phases decreases, which results into the initial increase in the transformation rate {\color{blue}\cite{Chaikin1994,Chaddah2008}}. However, the suppression of the relaxation rate on further reduction of temperature can be understood by the interplay between kinetic arrest and the free energy landscape {\color{blue}\cite{Chaddah2008,Pal2021}}.
\begin{figure}[t]
\centering
\includegraphics[scale=0.32]{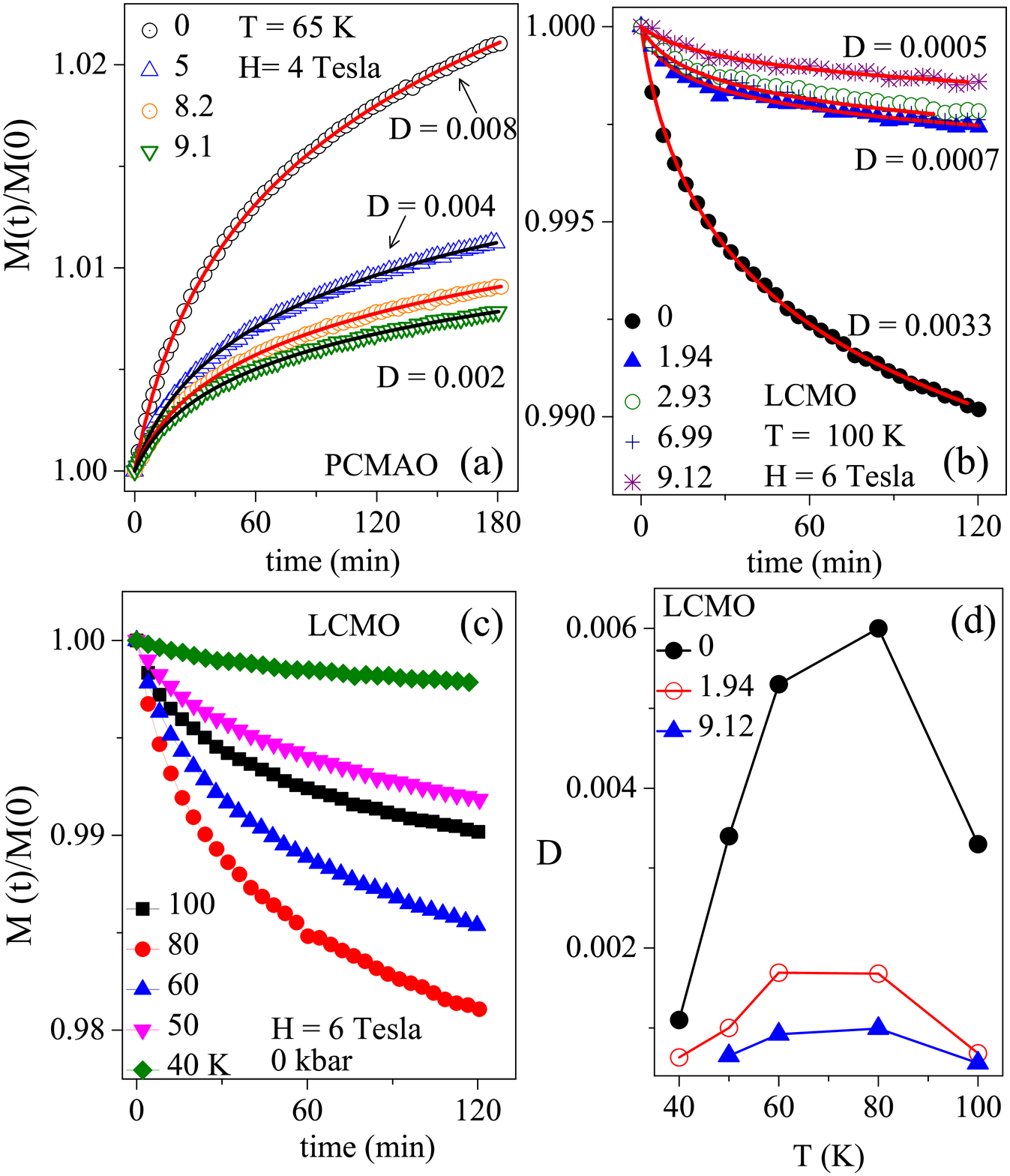}
\caption{Variation of M with time (a) of PCMAO at T = 65 K (b) of LCMO at 40 K measured in the FC protocol. The measurement fields are, H = 4 and 6 Tesla for PCMAO and LCMO respectively. Solid lines are fitted curve using eq. 2. D is the rate constant. The legends are the P values in the unit of kbar (c) the relaxation of M at various temperature across the thermal hysteresis at H = 6 Tesla and P = 0 in the FC protocol. The legends indicate the measurement temperatures. The curves are fitted using eqn. {\color{blue}2} and temperature dependence of D is shown in (d). It also presents the D(T) at higher P. The solid lines in (d) are guide to eye. } 
\end{figure}

\begin{figure}[t]
\centering
\hspace{-1cm}
\vspace{1cm}
\includegraphics[scale=0.36]{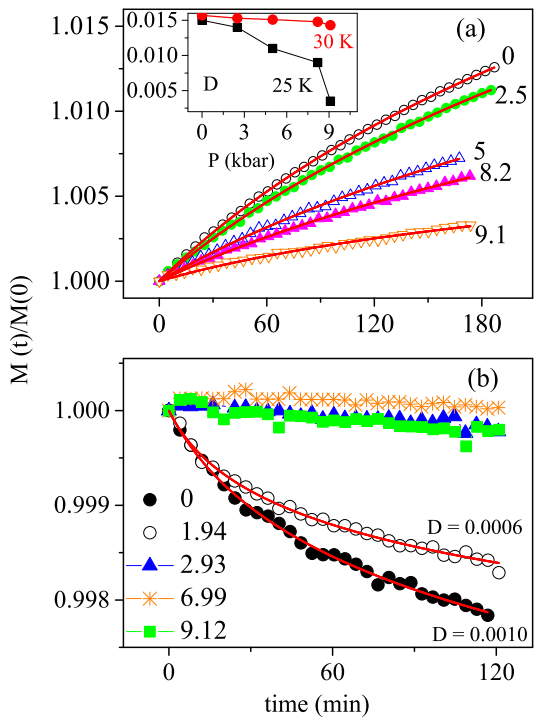}
\vspace{-1 cm}
\caption{Relaxation in the magnetic glass state in (a) PCMAO measured at T = 25 K and H = 4 Tesla and (b) LCMO measured at T = 40 K and H = 6 Tesla. Solid lines are the fitted curves using eqn. 2. The inset of (a) shows the D values at different P at T = 25 and 30 K in PCMAO obtained from fitting. The solid line in the inset is guide to eye. For LCMO, the curves at P = 0 and 1.94 kbar are fitted using same equation.} 
\end{figure}
The relaxation curves can be satisfactorily described by logarithmic dependence of M on time, which is given by:
\begin{equation}
M (t) = M(0) [1+ D log (1 + t/\tau)]
\end{equation}
Here, M(0) is the magnetization at time $t$ = 0, and $D$ is known as the rate constant which defines the relaxation time {\color{blue}\cite{Amir2009,Kustov2010, AB2011,Sudip2017,Sudip2019,Sudip2022}}. Such logarithmic function has been successfully used earlier to investigate the metastable behavior of the nonequilibrium state in spin glasses, structural glasses, magnetic glass etc.,{\color{blue}\cite{AB2011, Kustov2010, Sudip2017, Sudip2019,Sudip2022}}. We have fitted all the time dependence of M data in Figs. {\color{blue}6(a)} and {\color{blue}6(b)} using eqn. {\color{blue}2}. We have shown the fitted curves as the solid lines and mentioned the $D$ values in the respective figures. As can be observed here, in both PCMAO and LCMO, $D$ decreases with increasing P which further confirm that higher pressure inhibits the growth dynamics of the low temperature phases in both the systems in the course of the FM-AFM first order transition. Similarly, we have fitted the relaxation curves at various temperatures which are shown in Fig. {\color{blue}6(c)} for ambient P and plotted the obtained $D$ values in Fig. {\color{blue}6(d)}(black filled circles). Also, we have measured the relaxation of M at higher pressure, but we have not shown here for conciseness. We have fitted those data by using eqn. {\color{blue}2} and plotted only the $D$ values at a few selected higher P in Fig. {\color{blue}6(d)}. The variation of $D$ with temperature at different P is qualitatively similar. But, quantitatively,it is evident that external  P suppresses the growth rate in the entire temperature range of thermal hysteresis.

\subsection{Effect of pressure on the transformation kinetics in the magnetic glass state: } 
In the magnetic glass state, the non-equilibrium high-T phase which persists because of the KA gradually transforms into the equilibrium phase with time. The rate of transformation depends on the H and T. In Figs. {\color{blue}7(a)} and {\color{blue}7(b)}, we have shown the change in M as a function of time at temperatures T = 25 and 40 K following field cooled (FC) protocol in PCMAO and LCMO respectively. To record the M($t$) curves, the sample is initially cooled from T = 320 K to the measurement temperatures at different P in presence of the cooling fields of H = 4 and 6 Tesla for PCMAO and LCMO respectively. After temperature becomes stable, M has been recorded as a function of time for the next few hours without changing the temperature and field. The thermal hysteresis at H = 4 and 6 Tesla in PCMAO and LCMO close at T = 30 and 56 K respectively (data are not shown here for conciseness). Therefore, the measurement temperatures lie below the closer of the thermal hysteresis and situate in the magnetic glass regime of the respective materials. The relaxation of M with time indicates the non-equilibrium nature of the low-T state which can not be described in the framework of the metastable supercooled high-T phase. In case of PCMAO, M increases with time at ambient P which indicates that the AFM phase is unstable at this temperature and it transforms to the equilibrium FM phase. As P increases, the increase in M with time i.e., the relaxation rate monotonically decreases. It implies that the transformation rate monotonically reduces with the increase in P. On the other hand, in case of LCMO, the M decreases with time. It means that here the FM phase is the unstable phase and it gradually transforms into the equilibrium AFM phase. In this case, as P increases, the relaxation rate strongly decreases and becomes negligible at higher P. Therefore, P stabilizes the phase coexisting state in both the systems because it suppresses the relaxation rate of the FM-AFM transformation. However, at these field and temperatures, the suppression of the relaxation rate by P is pronounced in LCMO, compared to PCMAO. The relaxation curves can be fitted by using eqn. {\color{blue}2}. We have mentioned the obtained $D$ values for PCMAO in the main panel of Fig. {\color{blue}7(a)}. In the inset, we have shown the the variation of $D$ with P at T = 30 and 25 K. At T = 30 K, P suppresses the relaxation and $D$ decreases with increase in P. However, the effect of P is more pronounced at T = 25 K, which indicates that P strongly suppresses the transformation kinetics at lower temperatures. $D$ values for LCMO at P = 0 and 1.94 kbar have been mentioned in the Fig {\color{blue}7(b)}. At higher P, there is no appreciable relaxation of M.

\section{Discussion:}
In case of PCMAO, P monotonically  increases the supercooling and superheating temperatures which indicate that P favours the FM phase. However, P does not significantly affect the [H$_K$, T$_K$] band. It implies that the temperature window, where the first order transition occurs before it is arrested, expands with increasing P. This should increase the equilibrium FM volume fraction at low temperature, say T = 5 K. However, the transformation rate from the high-T AFM to the low-T FM phase across the thermal hysteresis is simultaneously suppressed by P [see Fig. {\color{blue}6(a)}]. Therefore, two competing mechanism acting together decides the phase fraction of the competing phases at any temperature.  The variation of  volume fraction of the FM phase at T = 5 K indicates that initially the second mechanism must be dominating over the first, so that the FM phase fraction initially decreases at low P but finally increases, when the first mechanism dominates. Note that, because of these two competing interactions, the change in the volume fraction with pressure is also small. On the other hand, in case of LCMO, $T^*$ reduces with increasing P and the $T_K$ band is unchanged. Hence, the temperature window where the first order transition from the high-T FM to low-T AFM phase occurs reduces with the increase in P. In addition, the nucleation and growth of the AFM phase is also suppressed by P. Therefore, both of these mechanisms together should result into the increase in the FM phase fraction at T = 5 K, which is observed in Fig. {\color{blue}2(b)}. Note that, in this case the change in the volume fraction is very large, which further support this scenario.

We would like to mention here that in PCMAO, the suppression of the transformation rate from the high-T AFM to low-T FM phase across the thermal hysteresis [see Fig. {\color{blue}6(a)}] by P is nontrivial. Because, as P increases the transition temperatures, the size of the critical nucleus of the FM phase at a particular temperature should be smaller at higher P {\color{blue}\cite{Chaikin1994,Chaddah}}. This should, in principle, increase the growth rate of the FM phase. We believe that this contradiction can be understood from the extremely different spin structure of the FM and AFM phases in PCMAO. In the course of the transition, the nucleus of a critical size (R$_C$) grows in the matrix of CE-AFM phase. The interface, where the two extremely different kind of spin structure co-exist, is expected to have considerable spin disorder that may cause hindrance to growth. Now, with increase in P, the R$_C$ decreases, the number of the FM nucleus as well as the interface will increase, which may suppress the growth rate at higher P. Note that, in LCMO, where AFM is also CE type and grows within the FM matrix, the transformation rate decreases with P. The same thing has been found at low temperatures i.e., in the magnetic glass state [see Figs. {\color{blue}7(a)} and {\color{blue}7(b)}]. In PCMAO, the transformation from the AFM to FM phase at T = 30 K is suppressed by P. Whereas, in LCMO the transformation from FM to AFM is hindered by P at T = 65 K. This reveals that higher P suppresses the transformation kinetics of the phase coexisting state in both systems. Basically,  the growth of one phase in the matrix of other phase appears to be suppressed at the interface of FM and CE-AFM. This conjecture needs to be verified by some other microscopic techniques. 

\begin{figure}[t]
\centering
\hspace{-1cm}
\vspace{1cm}
\includegraphics[scale=0.36]{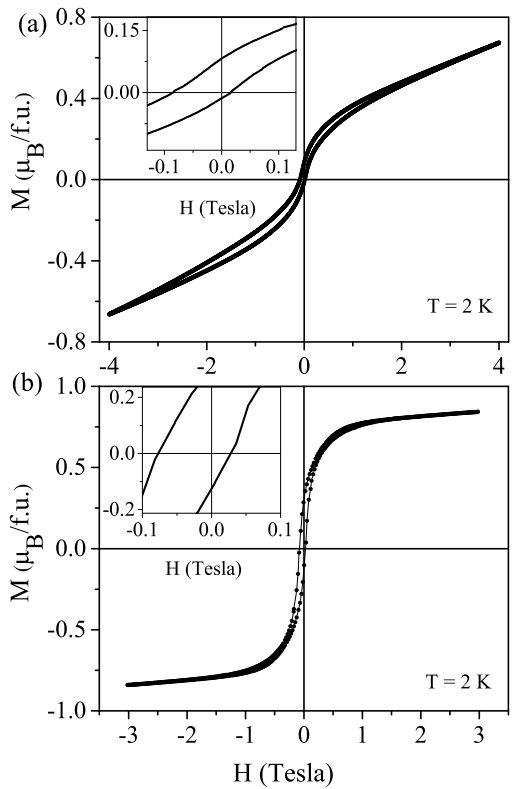}
\vspace{-1 cm}
\caption{(a) Isothermal M-H loop of PCMAO at T = 2 K. The sample has been cooled at H = 2 Tesla, and the hysteresis loop has been measured between $\pm$ 4 Tesla (b)  Isothermal M-H loop of LCMO at T = 2 K. The sample has been cooled at H = 1 Tesla, and the hysteresis loop has been measured between $\pm$ 3 Tesla. In the insets, we have shown the close view of the loop near origin.} 
\end{figure}
Nonetheless, the importance of the AFM-FM interface is evident from the observation of significant horizontal shift of the field cooled M-H loop in the phase coexistence region of both PCMAO and LCMO. Such shift in the M-H loop is popularly known as the exchange bias, and occurs due to pinning of the surface spins at the interface {\color{blue}\cite{EB1999}}. In Fig. {\color{blue}8(a)}, we have shown isothermal variation of magnetization with field (M-H curves) of PCMAO recorded after the samples has been cooled down to T = 2 K in presence of H = 2 Tesla, and the field cycle has been performed between $\pm$4 Tesla. Similarly, Fig. {\color{blue}8(b)} presents the field cooled M-H curve of LCMO at T = 2 K where the sample has been cooled at H = 1 Tesla, and the loop has been recorded between $\pm$3 Tesla. In the inset of the figures, we have shown the close view of the hysteresis loop near origin, which shows that the loop is not symmetric with respect to origin along the field axis, rather it is shifted towards the negative field axis. The exchange bias defined as $H_{EB}$ = $|(H_{C1}+H_{C2})|$/2, where $H_{C1}$ and $H_{C2}$ are the coercive fields along the positive and negative field axes respectively. In case of PCMAO and LCMO, the $H_{EB}$ has been found to be around $H_{EB}$ $\sim$ 350 and 245 Oe respectively. Such shift in the hysteresis loop arises due to coexistence of FM and AFM volume fractions in close proximity with each other.

\section{Summary:}   
In brief, we have studied the effect of simultaneously applied hydrostatic pressure and magnetic field on the phase coexistence in two well known perovskite manganites Pr$_{0.5}$Ca$_{0.5}$Mn$_{0.975}$Al$_{0.025}$O$_{3}$ and La$_{0.5}$Ca$_{0.5}$MnO$_3$. These two samples undergo first order phase transition between AFM and FM phases. In addition to the phase coexistence across the thermal hysteresis associated with conventional first order transition, the FM and AFM phases coexist even below the thermal hysteresis due to the phenomenon of kinetic arrest. During cooling, PCMAO undergoes transition from  high-T AFM to low-T FM phase. External P increases the supercooling and superheating temperatures, however, the kinetic arrest band is not significantly affected. The nucleation and growth rate of the FM phase from the AFM across the transition during cooling reduces at higher pressure. All these effects in combination results into a non-monotonic variation of the volume fraction of FM-metal and AFM-insulator phases at low temperature. At low temperature, i.e. below the thermal hysteresis, the AFM phase gradually transforms into the FM phase, and the transformation rate is suppressed at higher pressure. On the hand, LCMO has opposite phase diagram and it undergoes  a transition from high-T FM phase to low-T AFM phase while cooling. Pressure lowers the supercooling and superheating limits and increases the FM to AFM transformation rate across the transition, i.e. favors the AFM phase. With increase in P, the FM volume fraction at T = 5 K rises monotonically. In addition, the FM phase gradually transforms to the AFM phase and the transformation rate from FM to AFM slows down with increase in the pressure. The coexistence of different lattice and spin structure at the interface may be a possible reason behind such similar effect on two contrasting systems. Observation of exchange bias in both PCMAO and LCMO under field condition indicates the coupling of the FM and AFM phase at the interface.\\

\end{document}